\documentclass[final,5p]{elsarticle}
\usepackage{amsmath}
\usepackage{amssymb}
\usepackage{graphicx}
\usepackage{hyperref}
\usepackage[all]{hypcap}
\usepackage{natbib}
\usepackage{subfigure}
\begin{document}
\begin{frontmatter}

\title{A Mathematical Approach to the Study of the United States Code}

\author{Michael J. Bommarito II}
\ead{mjbommar@umich.edu}
\address{Center for the Study of Complex Systems, University of Michigan, Ann Arbor}
\address{Department of Political Science, University of Michigan, Ann Arbor}
\address{Department of Financial Engineering, University of Michigan, Ann Arbor}

\author{Daniel M. Katz}
\ead{dmartink@umich.edu}
\address{Center for the Study of Complex Systems, University of Michigan, Ann Arbor}
\address{Department of Political Science, University of Michigan, Ann Arbor}
\address{University of Michigan Law School}

\date{\today}

\begin{abstract}
The \textit{United States Code} (Code) is a document containing over 22 million words that represents a large and important source of Federal statutory law.  Scholars and policy advocates often discuss the direction and magnitude of changes in various aspects of the Code.  However, few have mathematically formalized the notions behind these discussions or directly measured the resulting representations.  This paper addresses the current state of the literature in two ways.  First, we formalize a representation of the \textit{United States Code} as the union of a hierarchical network and a citation network over vertices containing the language of the Code.  This representation reflects the fact that the Code is a hierarchically organized document containing language and explicit citations between provisions.  Second, we use this formalization to measure aspects of the Code as codified in October 2008, November 2009, and March 2010.  These measurements allow for a characterization of the actual changes in the Code over time.  Our findings indicate that in the recent past, the Code has grown in its amount of structure, interdependence, and language.
\end{abstract}

\end{frontmatter}

\section{Formalizing the Code}
The \textit{United States Code} (Code) is a document containing over 22 million words that represents a large and important source of Federal statutory law.  The Code is a concise and conveniently organized compilation of all ratified legislation and treaties, and is often the first source of information for lawyers, judges, and legal academics.\footnote{The complete set of all ratified legislation and treaties is known as the \textit{Statutes at Large}.  As a legal technicality, the Code is only \textit{prima facie} evidence of Federal law.  In the event of a discrepancy, the \textit{Statutes at Large} are the final authority.  Furthermore, additional sources such as the \textit{Code of Federal Regulations} contains clarifications issued by other Federal agencies or bodies.}  The Code is compiled through a process known as codification, which is carried out by the \textit{Office of the Law Revision Counsel} (LRC), an organization within the U.S. House of Representatives.  The LRC's goal in this codification process is to transform the incremental and chronological \textit{Statutes at Large} into the Code, a current snapshot of the law that is organized into hierarchical categories.\footnote{2 U.S.C. \S 285- \S285g outlines the purpose, policy and functions of the Office of Law Revision Counsel.}  

This hierarchical organization is an important qualitative feature of the Code.  At the first level of the hierarchy, the Code is divided into 49 titles that represent the broadest categories of law.  Well-known titles include the Tax Code, formally known as \textit{Title 26 - Internal Revenue Code}, \textit{Title 20 - Education}, and \textit{Title 18 - Crimes and Criminal Punishment}.  Each title is also hierarchically subdivided into some combination of subtitles, chapters, subchapters, parts, subparts, sections, subsections, paragraphs, subparagraphs, clauses, or subclauses.  Out of these vertex types, only sections, subsections, paragraphs, subparagraphs, clauses and subclauses can contain the actual substantive legal text.

The text within these vertices can also contain explicit citations to other portions of the Code.  These citations may be used to reference definitions, highlight qualifying conditions, or point to well-understood processes.  It is critical to recognize that these citations are not restricted by the organizational hierarchy.  For example, sections within \textit{Title 26}, the Tax Code, can and do contain citations to \textit{Title 18}, the Criminal Code.  Thus, though the LRC attempts to organize the Code into a cleanly divided hierarchical network, there is also a citation network embedded within the Code that is not constrained by this hierarchy.

Based on the characterization above, we can formulate a mathematical representation of the Code as a graph $\mathcal{G} = (V,E)$ with an associated ``text'' function $T(v)$.  $V$ is the set of vertices composed of all titles, subtitles, chapters, subchapters, parts, subparts, sections, subsections, paragraphs, subparagraphs, clauses, and subclauses.  These vertices can also be divided into two subsets: (1) vertices that do contain text, written $V^T$, and (2) vertices that do not contain text, written $V^N$.  For vertices $v \in V^T$ that do contain text, the associated ``text'' function $T(v)$ returns an ordered tuple containing the tokens within the text of vertex $v$.  For vertices $v \in V^N$ that do not contain text, $T(v) = \emptyset$.  The set of edges $E$ can likewise be divided into two subsets: (1) edges that encode the hierarchical organization of the Code, written $E^H$, and (2) edges that record the citation network within the Code, written $E^C$.  For convenience, we can then write the edge-induced subgraphs $\mathcal{G}^H$ and $\mathcal{G}^C$ that represent the hierarchical network and citation network of the Code respectively.  

For the remainder of this study, we fold the subtree under each section vertex back into its respective section vertex.  For example, the text and citations from 26 U.S.C. \S 501(c)(3) are merged up into 26 U.S.C. \S 501. While this choice trades off some amount of detail in order to compare the properties of $T(v)$ across snapshots, we believe there are several compelling justifications supporting this choice. By focusing on sections, we ensure that all leaf vertices of the hierarchy are of the same type.  This makes a number of network calculations much simpler and easier to interpret than otherwise.  Furthermore, unlike other vertices in $V^T$, sections are the only type of vertex that is guaranteed to contain complete grammatical units.  This makes section vertices the natural unit of analysis for any statements regarding the language within the Code.

Though our attention throughout the remainder of the paper is on the Code as a mathematical object, it is worthwhile to note that there are many important objects that exhibit characteristics which are qualitatively similar to the Code.  For example, Internet web pages are hierarchically structured by IANA country code, domain, subdomain, and directory structures.  These web pages also contain large amounts of language and explicit interdependence in the form of hyperlinks.  Therefore, the analysis carried out in this paper could also be applied to web pages on the Internet or any similar document.  In summary, our representation of the Code can therefore be more generally described as a formalization of a document with hierarchical \textbf{structure}, explicit \textbf{interdependence}, and a significant amount of \textbf{language}.

\section{Measuring the Code}
We can measure aspects of this representation of the Code by constructing it from empirical data.  To do so, we have obtained XML snapshots of the Code at three points in time: October 2008, November 2009, and March 2010.  This data was provided by the Cornell Legal Information Institute (\cite{LII}).  It is important to understand that the rate of legislation and treaty-making exceeds the LRC's rate of codification.  Furthermore, the LRC codification schedule is based on titles, not on the chronology of the \textit{Statues at Large}.  As a result, we cannot make compare the rate of growth of different sections or titles of the Code with this data.  We can, however, make statements with respect to aggregate changes in the Code.

\begin{figure}[htb]
	\centering
	\subfigure[$\mathcal{G}^H$]{\includegraphics[width=7cm]{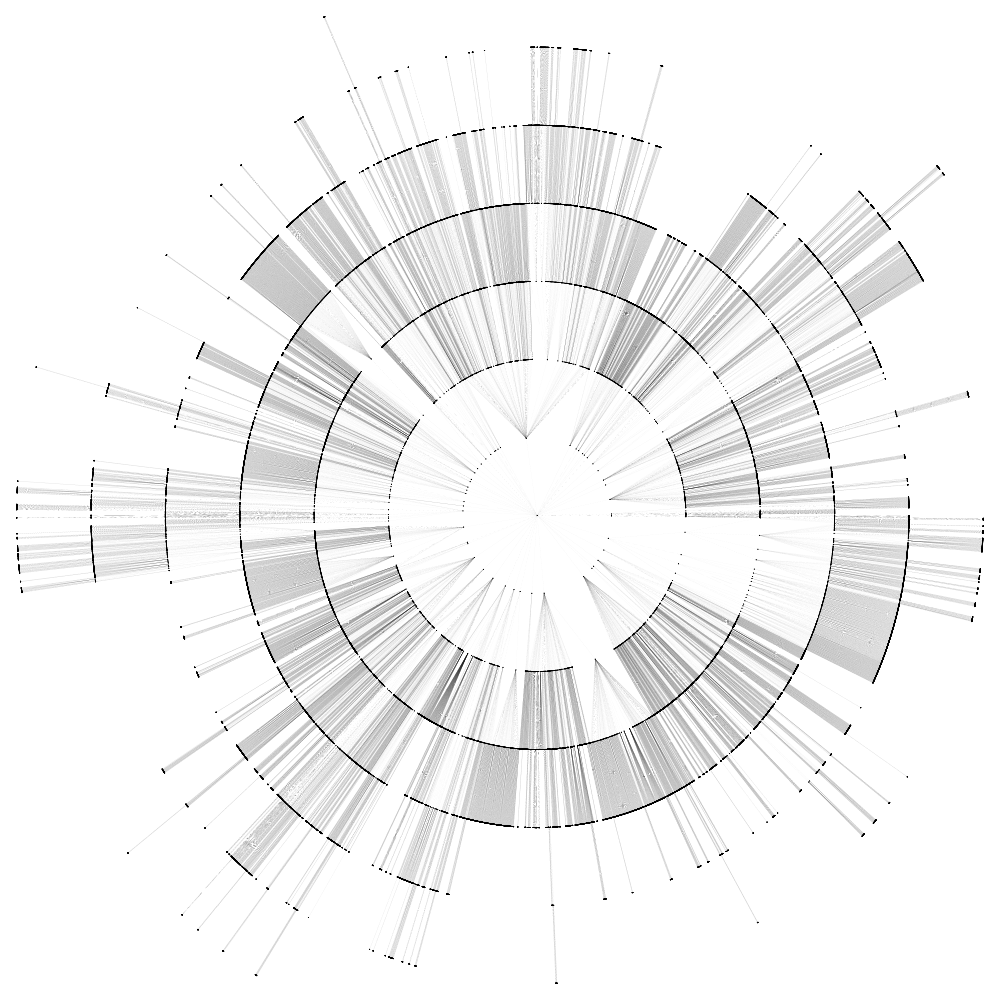}}
	\subfigure[$\mathcal{G}$]{\includegraphics[width=7cm]{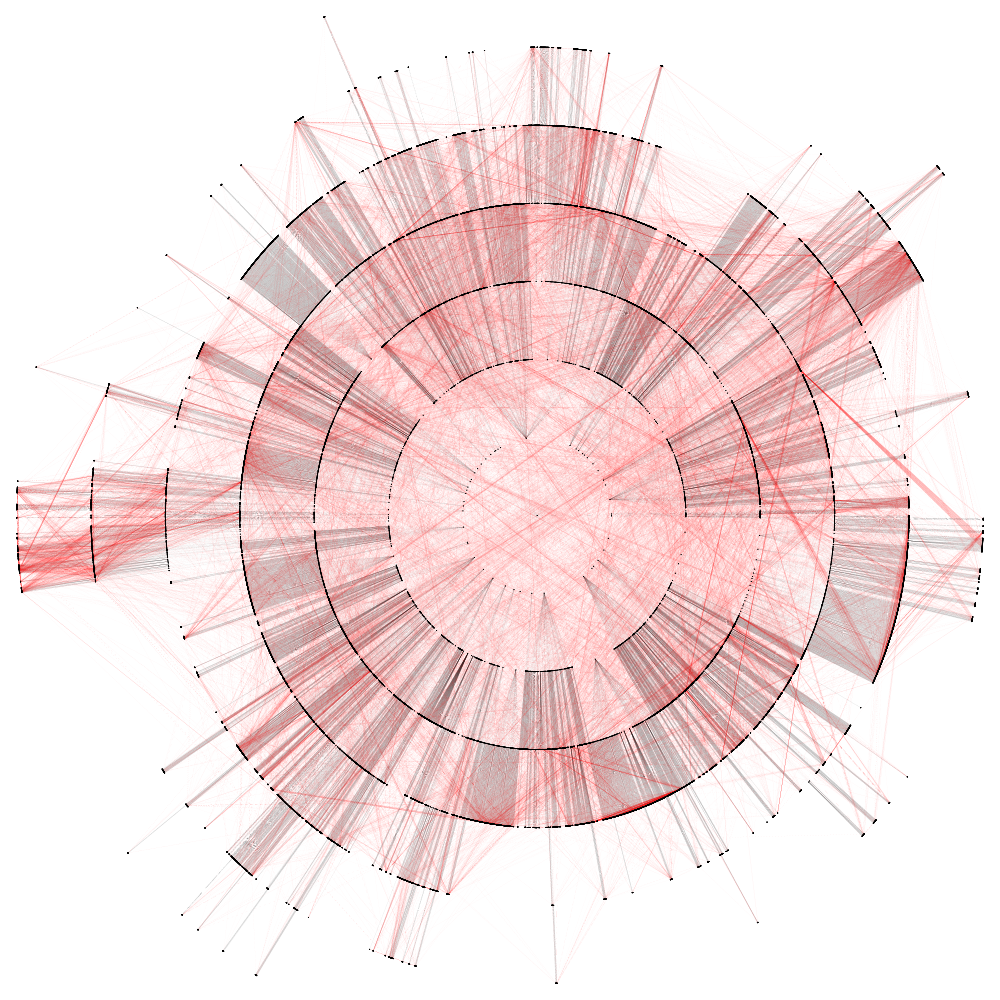}}
	\caption{Network Visualization of the Code, Oct. 2008. Top: Hierarchical network, Reingold-Tilford circular layout (\cite{Reingold1981}). Bottom: Hierarchical network with citation network overlay in red.}
	\label{fig:snapshotexample}
\end{figure}

Figure \ref{fig:snapshotexample} offers two visualizations of the snapshot of the Code as of October 2008.  In panel (a), the hierarchical network $\mathcal{G}^H$ is shown branching out from the rooting node.  In panel (b), the section-to-section citation network is imposed in red onto the hierarchical network in (a).  Note that all citations are from one leaf of the hierarchical network to another leaf of the hierarchical network.  This is a consequence of our above choice to combine any subsections and their text into their respective section vertices.

\begin{table}[htb]
	\centering
	\small
	\begin{tabular}{|c|c|c|c|c|}
		\hline
		\textbf{Date} & $|V(\mathcal{G})|$ & $|E(\mathcal{G})|$ & $|V(\mathcal{G}^C)|$ & $|E(\mathcal{G}^C)|$\\\hline
		Oct 2008 & 57947 & 140154 & 33503 & 82208\\\hline
		Nov 2009 & 59684 & 144758 & 34473 & 85075\\\hline
		Mar 2010 & 59988 & 145908 & 34674 & 85921\\\hline
	\end{tabular}
	\caption{Summary of Code snapshots.}
	\label{table:snapshottable}
\end{table}

Figure \ref{fig:snapshotexample} clearly demonstrates that the Code is a large object.  Table \ref{table:snapshottable} shows a summary of  the size of the Code and its edge-induced citation network for each snapshot.  We do not need to write $|V(\mathcal{G}^H)|$ and $|E(\mathcal{G}^H)|$ separately, since $|V(\mathcal{G}^H)| = |V(\mathcal{G})|$ and $|V(\mathcal{G}^H)| = |E(\mathcal{G}^H)| + 1$ by construction.  

This table gives us the first opportunity to examine changes in the Code.  In assessing such changes, it is important to note that the lag associated with the LRC codification schedule complicates the relationship between passage of individuals pieces of legislation and their subsequent reflection in the Code.  Despite this limitation, we can still consider aggregate net changes in the Code over our window of data.

We first focus on the hierarchical structure of the Code as represented by $\mathcal{G}^H$.  Between October 2008 and November 2009, approximately 4.29 vertices were added to the Code each day, and between November 2009 and March 2010, approximately 2.71 vertices were added to the Code each day.  It is important to note that vertex creation can be the product of (1) reorganization of existing hierarchy and its embedded language or (2) the codification of new legislation.  In the first case, the total amount of language in the Code is often relatively unchanged.  In the second case, however, new language from legislation must be introduced and the total amount of language in the Code increases.\footnote{In some cases, new legislation replaces old legislation by repealing existing language and structure in the Code and replacing it with new language and structure.  The ``vertex creation'' rates above are thus more accurately labeled as ``net vertex creation'' rates.}  Regardless of which dynamic is responsible, these rates show that the Code has experienced robust growth in its hierarchical structure in recent years.

\begin{figure}[htb]
	\centering
	\includegraphics[width=8cm]{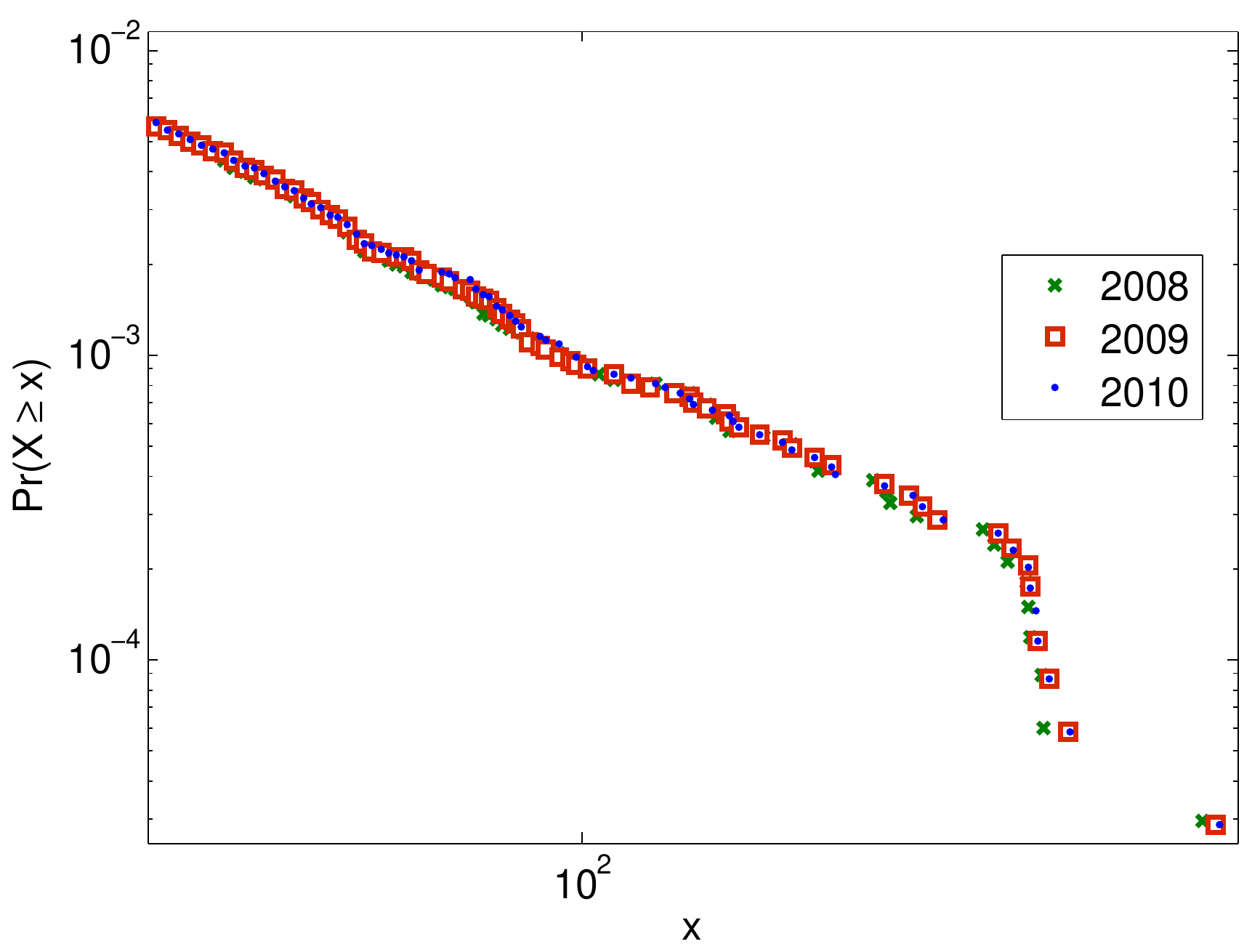}
	\caption{Log-Log Degree Distribution for $\mathcal{G}^C$ over Time.  Plotted with Matlab code from \cite{Clauset2009}.}
	\label{fig:citation_degree}
\end{figure}

In order to assess changes in the interdependence of the Code, we now examine $\mathcal{G}^C$ and its corresponding degree distribution.\footnote{Readers interested in properties of the degree distribution alone should refer to \cite{Bommarito2009}.}  Table \ref{table:snapshottable} shows that both $|V(\mathcal{G}^C)|$ and $|E(\mathcal{G}^C)|$ have been growing at a rapid pace.  Between October 2008 and November 2009, approximately 2.4 sections entered the citation network and approximately 7.08 new citations were added per day.\footnote{Sections may ``enter'' the citation network in two ways: (1) by being newly created and containing citations in their original form or (2) by having previously existed but either receiving their first citation or being amended to cite another section for the first time.}  Between November 2009 and March 2010, approximately 1.79 sections entered the citation network per day and approximately 7.55 new citations were added per day.  These numbers clearly indicate that the recent trend is positive.

Figure \ref{fig:citation_degree} shows the region of the log-log degree distribution that exhibits change across these snapshots.  Though a power-law distribution is soundly rejected for all three snapshots, the standard deviation and skewness of the degree distributions increase from 2008 to 2009 and 2009 to 2010 (\cite{Clauset2009}).  Furthermore, the maximum degree also increased from 2008 to 2009 and 2009 to 2010.  Based on these simple measurements of the Code, it is clear that the Code is growing both in the size of its hierarchical structure and in the amount of conceptual interdependence.

We next turn our attention to the language contained within the Code.  Though our mathematical representation of the Code relegates this language to the function $T$, we should remember that language and its organization are the components most commonly invoked in the popular discourse surrounding the Code. The first step in this analysis is to examine the change in the amount of language in the Code over time.  Table \ref{table:word_dist} allows us to assess these changes by comparing the language contained within the snapshots of the Code over time.  The numbers given in this table represent the average size of each section, i.e., $|T(v)|$, and the total size of the Code as measured by ``words.''\footnote{Here, ``words'' correspond to those strings tokenized by the Penn Treebank and are not necessarily restricted to words that might be found in a dictionary (\cite{Marcus1994}).}  

Just as Table \ref{table:snapshottable} above shows that structure of the Code is growing, Table \ref{table:word_dist} shows that the amount of language in the Code is growing.  This growth is at a rate of approximately 2,730 words per day between October 2008 and November 2009 and 2,706 words per day between November 2009 and March 2010.  Furthermore, since the average word count per section is also increasing over these time periods, the rate of increase in language surpasses the rate of growth in the number of sections.  In other words, the language in the Code is growing both through the addition of new sections containing new text and through the lengthening of existing sections.  

\begin{table}[htb]
	\centering
	\small
	\begin{tabular}{|c|c|c|c|}
		\hline
		\textbf{Year} & \textbf{Word Count per Section} & \textbf{Total Word Count}\\\hline
		2008 & 468.16 & 22,823,405\\\hline
		2009 & 476.03 & 23,919,248\\\hline
		2010 & 479.85 & 24,224,985\\\hline
	\end{tabular}
	\caption{Word Count per Section and Total Word Count over Time.}
	\label{table:word_dist}
\end{table}

\begin{figure}[htb]
	\centering
	\includegraphics[width=7cm]{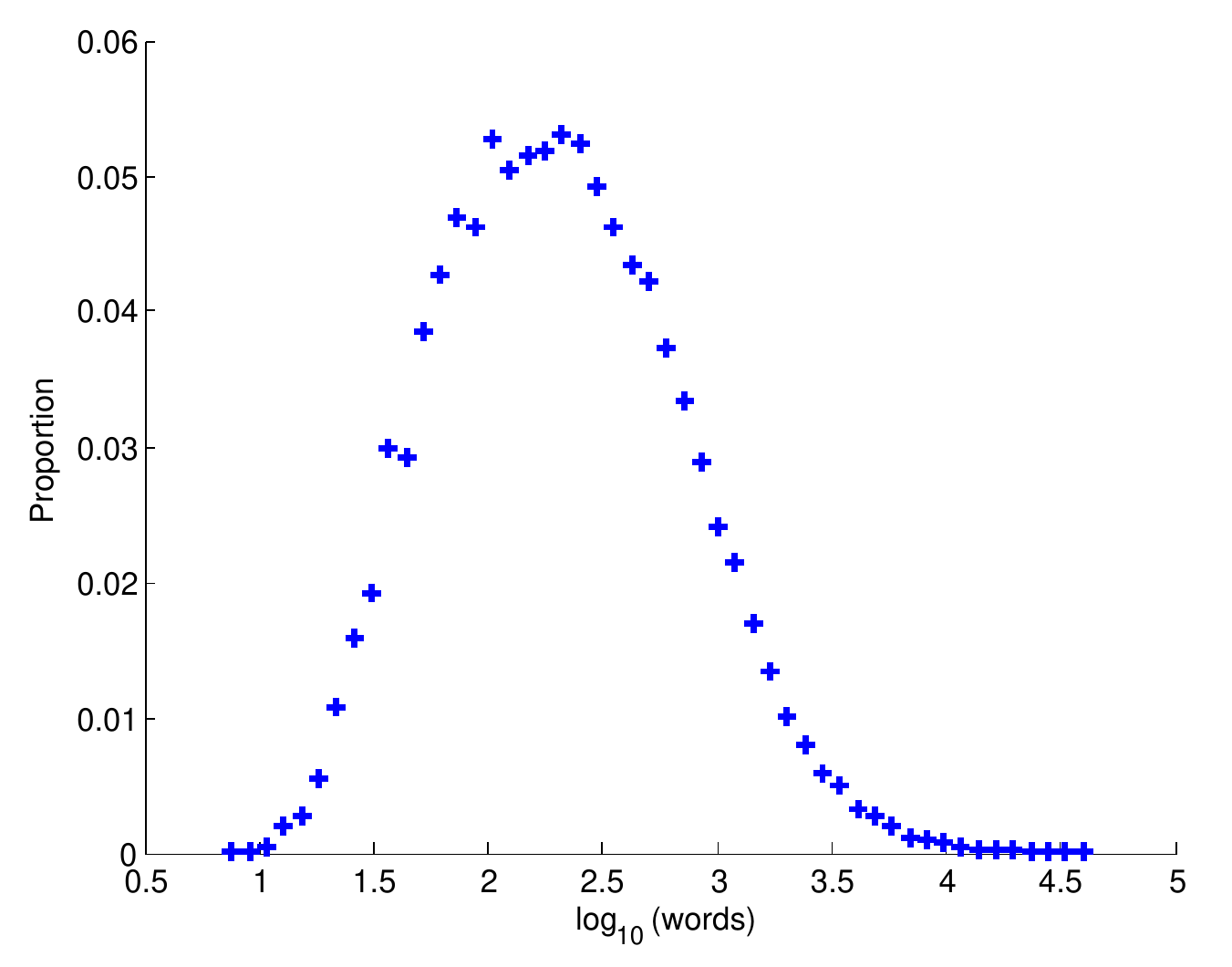}
	\caption{Distribution of Section Sizes by $log_{10}$ Word Count in October 2008.}
	\label{fig:word_dist}
\end{figure}

Underlying each row in Table \ref{table:word_dist} is an entire distribution of word counts per section.  Examining the change in these distributions over time can better explain the changes evidenced in the Table.  Figure \ref{fig:word_dist} shows the distribution of word count per section for the Code as of October 2008.  This figure shows that section sizes are fairly normally distributed around the mean size of 468.16 words per section with a standard deviation of 1,031.2.

\begin{figure}[htb]
	\centering
	\includegraphics[width=7cm]{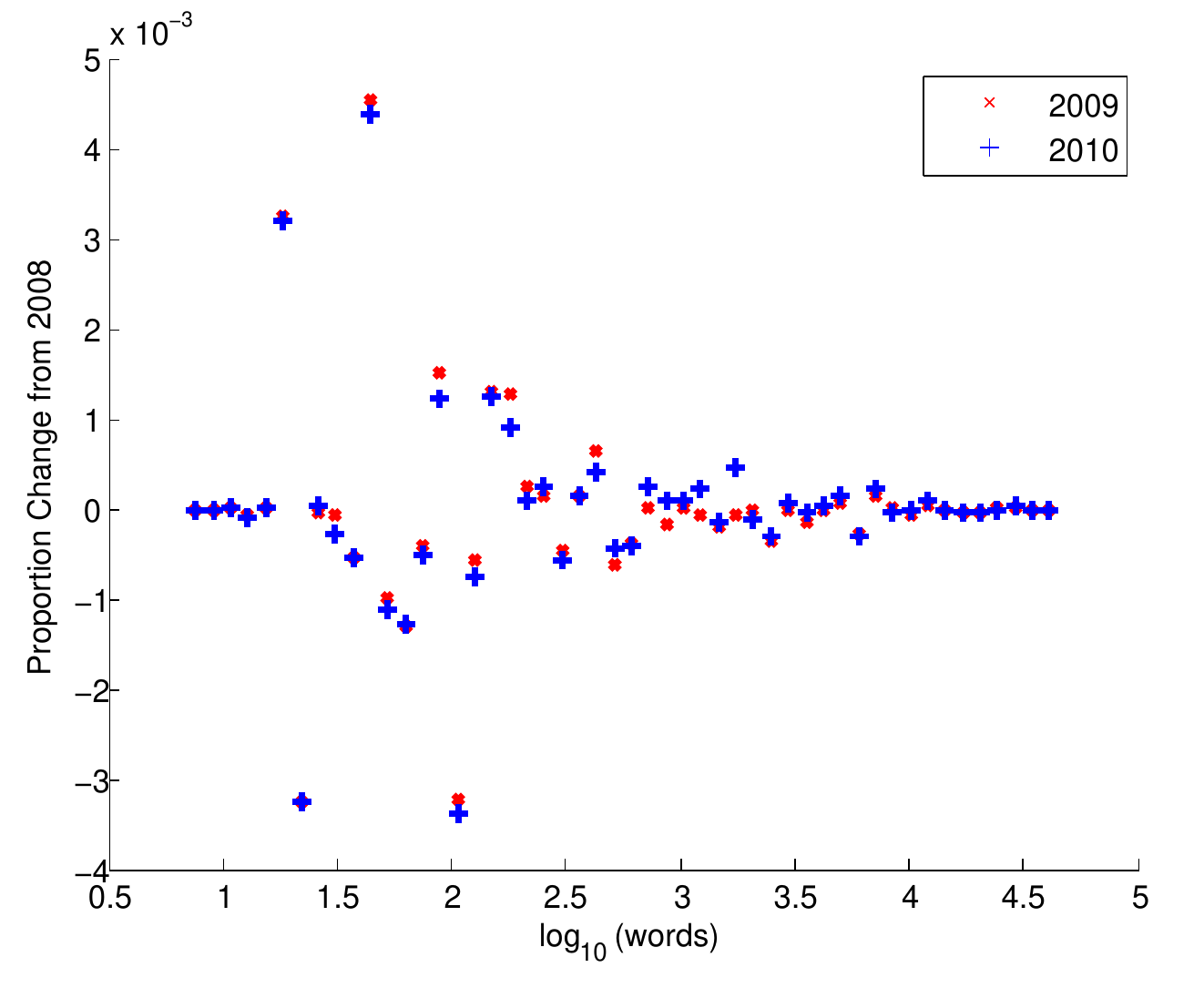}
	\caption{Changes in the Distribution of Section Sizes by $log_{10}$ Word Count from 2008 over Time.}
	\label{fig:word_change_dist}
\end{figure}

The differences between the distributions are shown in Figure \ref{fig:word_change_dist}.  The circles and pluses show the change in the section size distribution from October 2008 to November 2009 and from October 2008 to March 2010 respectively.  This figure shows that the changes are relatively small and primarily contained in the bulk of the distribution in sections between 10 and 1,000 words.  Furthermore, the overall trend in changes indicates an increasing number of sections at sizes between $10^2$ and $10^{3.5}$.  The maximum section size did increase from 43,183 words in October 2008 to 44,962 words in November 2009.  These observations imply that the average section size is increasing due to a larger number of sections in this region, not because of a drastic change in the sizes of sections contained within the Code.

Language can change in ways that are not captured by word counts.  One important factor that is not captured by the number of words alone is the \textit{distribution} of these words.  For example, a section with a word count of 100 may contain either a single word repeated over and over or it may contain 100 different words each used once.  In order to capture differences such as these, we next investigate the Shannon entropy of the word distribution for each section of the Code (\cite{Shannon1948}).  Entropy values will allow us to distinguish between the above cases, with the former example having much lower entropy than the latter example.  

In many ways, the entropy of the section word distribution can be related to some of the characterizations of complexity offered by legal scholars (\cite{Long1987} \cite{Schuck1992} \cite{Rook1993}).  If we assume that words map to distinct concepts, then this measure corresponds to the uncertainty or variance in the concept distribution.  Concept variance is important because, all else equal, it is more difficult for an individual to understand a set of concepts with high variance than one comprised of homogeneous material.  However, there is a significant difference between complexity and difficulty and we should be careful to explain entropy as ``concept variance'' (\cite{Page2008}).

In order to calculate the section entropy for a section $v \in V^T$, we need to calculate 
\begin{align*}
	W =& \{w : w \in T(v)\}\\
	H =& -\sum_{w \in W} p(w) \log_2 p(w)
\end{align*}
where $p(w)$ is the proportion of tokens in $T(v)$ that are $w$.  By calculating $H$ for all $v \in V^T$, we can observe the actual distribution of section entropy.

\begin{figure}[htb]
	\centering
	\includegraphics[width=7cm]{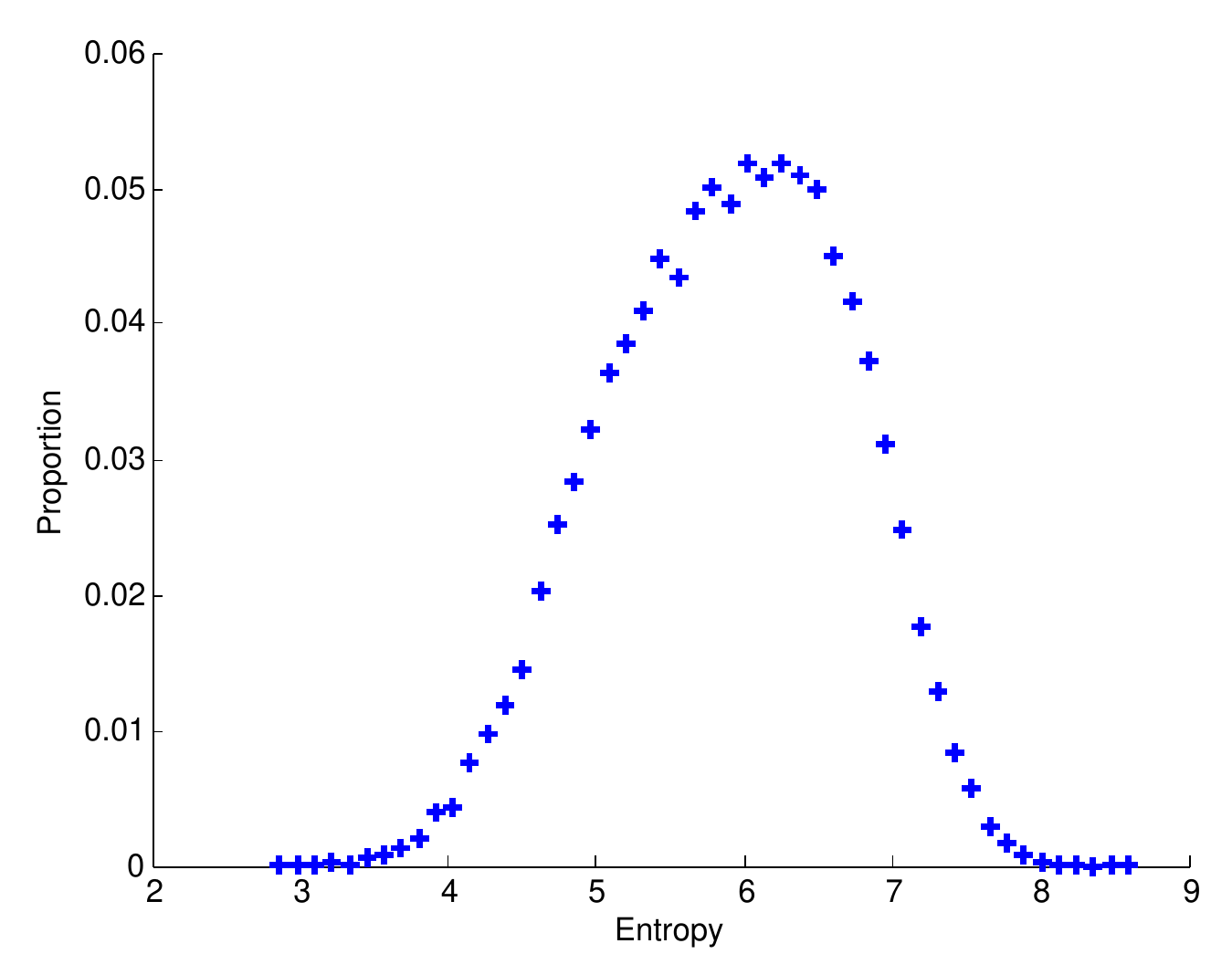}
	\caption{Distribution of Section Entropy in October 2008.}
	\label{fig:entropy_dist}
\end{figure}

Figure \ref{fig:entropy_dist} shows the distribution of section entropy in October 2008.  The distribution is relatively normal with a right lean.  The mean section entropy is 5.89 with a standard deviation of 0.81, and the maximum section entropy is 8.65. As a comparison, the first two paragraphs of this paper have an entropy of approximately 6.7.  

\begin{figure}[htb]
	\centering
	\includegraphics[width=7cm]{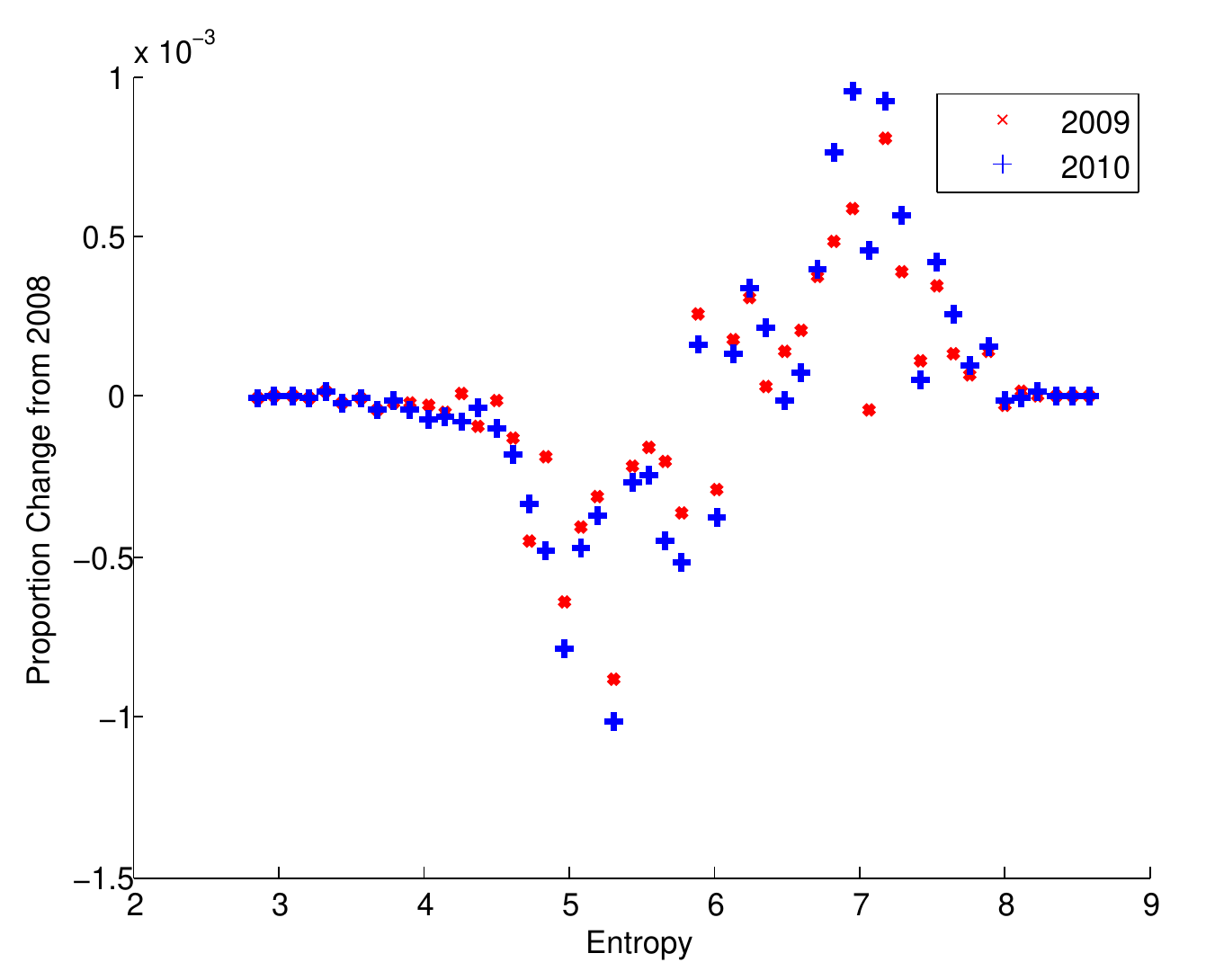}
	\caption{Changes in the Distribution of Section Entropy from 2008 over Time.}
	\label{fig:entropy_change_dist}
\end{figure}

Figure \ref{fig:entropy_change_dist} plots the distribution of changes in the section entropy distributions over time.  The circles and pluses correspond to the change in the section size distribution from October 2008 to November 2009 and from October 2008 to March 2010 respectively.  This figure shows a strong trend away from the October 2008 snapshot for both November 2009 and March 2010.  The proportion of sections with entropy between 4.5 and 6 has decreased nearly symmetrically with the increase in sections with entropy between 6 and 8.  Furthermore, the trend is more pronounced in the March 2010 snapshot than in the November 2009 snapshot.  These observations imply that there is a shift from less conceptual variance per section to more conceptual variance per section, implying an increase in ``legal complexity'' over our window of observation.

\section{Conclusion \& Future Work}
In this paper, we offer a mathematical formalization of the \textit{United States Code} that incorporates its important qualitative features including hierarchical structure, explicit interdependence, and language.  We use this representation to measure the changes in the Code in the period between October 2008 and March 2010.  These measurements imply that in the recent past the Code has become larger both in its hierarchical structure and its language.  Furthermore, the sections of the Code have become more explicitly interdependent.

In future work, we hope to obtain snapshots of the Code over a longer period of time.  By obtaining data for at least one LRC codification cycle, we could evaluate both aggregate changes over longer windows and changes in specific sections or titles of the Code.  This would allow us to directly evaluate claims made by both policy advocates and scholars.  Furthermore, we hope to more seriously address the concept of ``legal complexity'' by developing a complexity measure based on this mathematical representation of the Code.

\section{Acknowledgments}
We would like to thank the Center for the Study of Complex Systems (CSCS) at the University of Michigan for a fruitful research environment.  This work was partially supported by an NSF IGERT fellowship through the Center for the Study of Complex Systems (CSCS) at the University of Michigan, Ann Arbor.  We would also like to thank Dave Shetland at the Cornell Legal Information Institute and Joel Slemrod, J.J. Prescott, and Abe Gong at the University of Michigan for their assistance and feedback.

\end{document}